\documentclass[useAMS,usenatbib,usegraphicx]{mn2e}

\pdfoutput=1 
\usepackage{amsmath}
\usepackage{color}

\newif\ifAMStwofonts
\AMStwofontstrue

\newcommand{\simlt}{\lower.5ex\hbox{$\; \buildrel < \over \sim \;$}}
\newcommand{\simgt}{\lower.5ex\hbox{$\; \buildrel > \over \sim \;$}}

\title[Dust properties of M82: Radial variations]
{Variations of the dust properties of M82 with galactocentric distance}
\author[S. Hutton, I. Ferreras and V. Yershov]
{Susan Hutton, Ignacio Ferreras\thanks{E-mail: i.ferreras@ucl.ac.uk}, 
Vladimir Yershov\\
Mullard Space Science Laboratory, University College London, 
Holmbury St Mary, Dorking, Surrey RH5 6NT\\
}

\begin{document}
\date{Accepted for publication in MNRAS, 11 June 2015}
\pagerange{\pageref{firstpage}--\pageref{lastpage}} \pubyear{2015}
\maketitle
\label{firstpage}

%%\newif\ifAMStwofonts
%%\AMStwofontstrue

\begin{abstract}
We use near ultraviolet and optical photometry to investigate the dust
properties in the nearby starburst galaxy M82. By combining imaging
from the Swift/UVOT instrument and optical data from the Sloan Digital
Sky Survey, we derive the extinction curve parameterized by the
standard $R_V$ factor, and the strength of the NUV 2175\,\AA\ feature
-- quantified by a parameter $B$ -- out to projected galactocentric
distances of 4\,kpc. Our analysis is robust against possible
degeneracies from the properties of the underlying stellar
populations. Both $B$ and $R_V$ correlate with galactocentric
distance, revealing a systematic trend of the dust properties. Our
results confirm previous findings that dust in M82 is better fit by a
Milky Way standard extinction curve (Hutton et al.), in contrast to a
Calzetti law. We find a strong correlation between $R_V$ and $B$,
towards a stronger NUV bump in regions with higher $R_V$, possibly
reflecting a distribution with larger dust grain sizes. The data we
use were taken before SN2014J, and therefore can be used to probe the
properties of the interstellar medium before the event. Our $R_V$
values around the position of the supernova are significantly higher
than recent measurements post-SN2014J ($R_V\!\approx\!1.4$). This
result is consistent with a significant change in the dust properties
after the supernova event, either from disruption of large grains or
from the contribution by an intrinsic circumstellar component.
Intrinsic variations among supernovae not accounted for could also
give rise to this mismatch.
\end{abstract} 


%%%%%%%%%%%%%%%%   Figure 1  %%%%%%%%%%%%%%%%%%%
%%%%%%%%%%%%%%%%%%%%%%%%%%%%%%%%%%%%%%%%%%%%%%%%
\begin{figure*}
\begin{center}
\includegraphics[width=8.7cm]{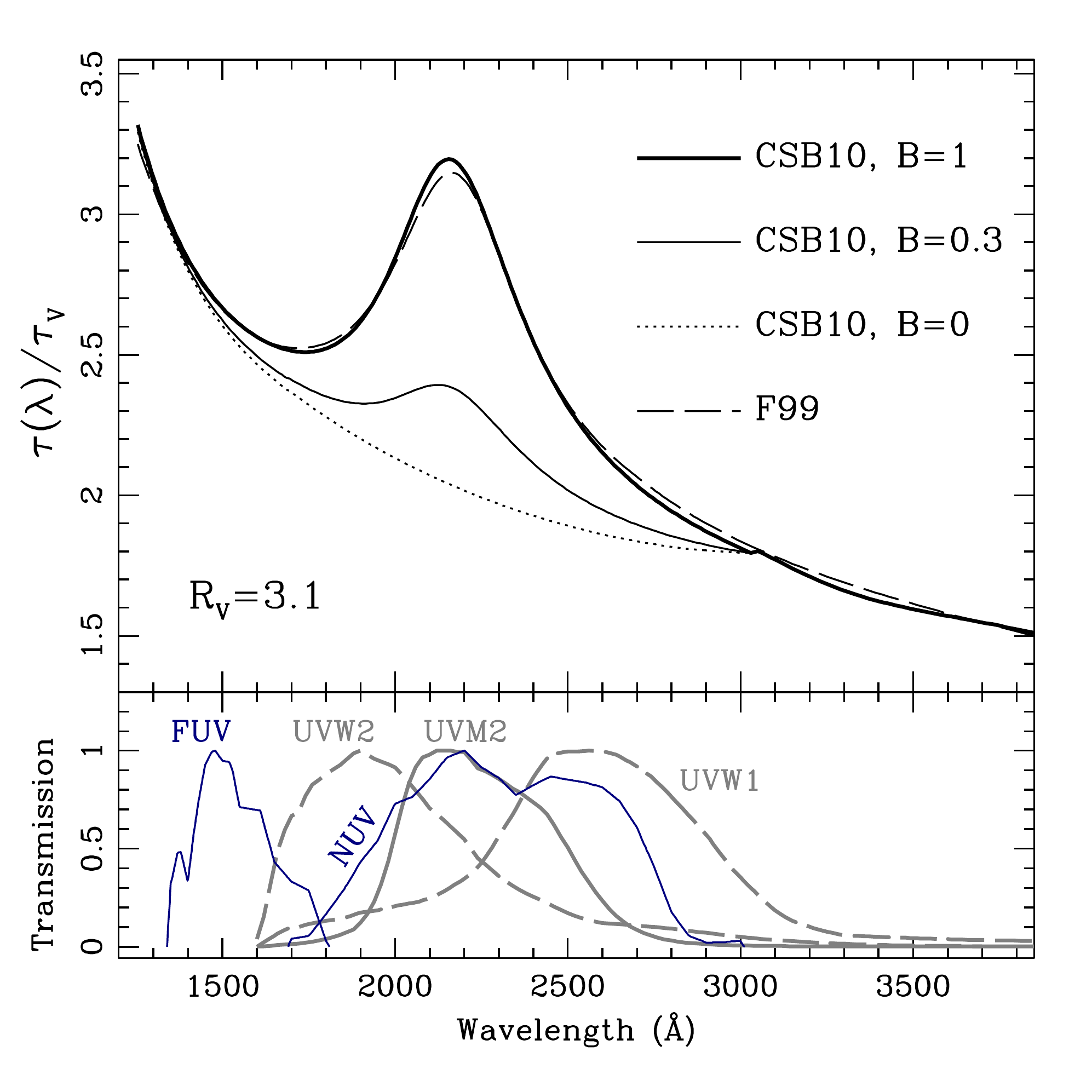}
\includegraphics[width=8.7cm]{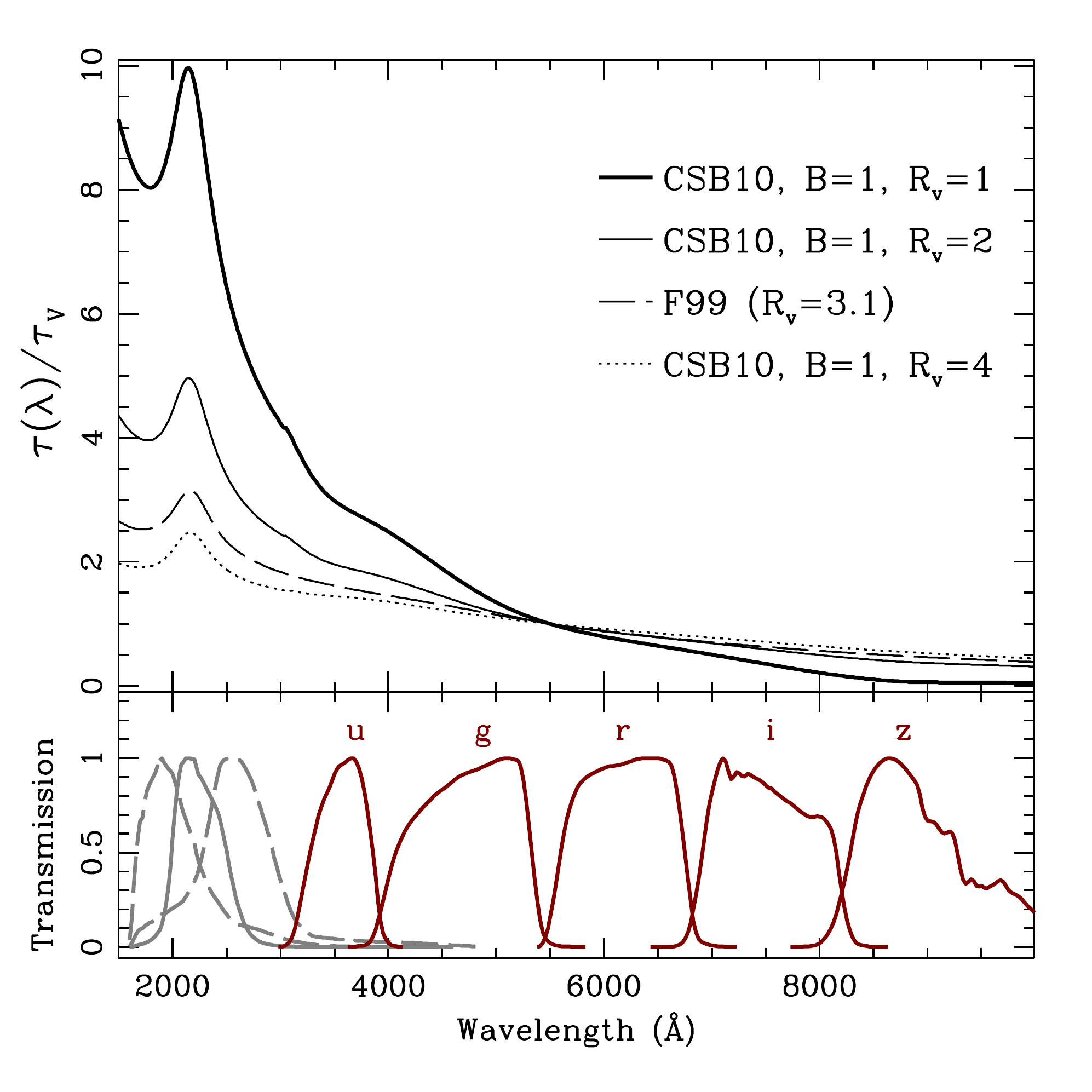}
\end{center}
\caption{{\sl Top:} Extinction law in the NUV ({\sl left}) and 
NUV+optical ({\sl right}) regions
from \citet[][CSB10]{CSB:10}. Several cases of $R_V$ and the NUV bump
parameter $B$, are shown, as labelled. The CSB10 models on the
left-hand panel all assume $R_V=3.1$. In both figures, the dashed line
shows for reference the Milky Way-standard extinction law
\citep[][F99]{Fitz:99}. {\sl Bottom:} In grey we show the response functions of
the NUV filters of Swift/UVOT (UVW2, UVM2 and UVW1).  The passbands of
GALEX (FUV and NUV) are also shown on the left-hand panel, and the
filter response curves of SDSS ($u,g,r,i,z$) are shown in the
right-hand panel. For each filter the effective throughput is divided
by the peak transmission.}
\label{fig:atten}
\end{figure*}
%%%%%%%%%%%%%%%%%%%%%%%%%%%%%%%%%%%%%%%%%%%%%%%%

%%%%%%%%%%%%%%%%%%%%%%%%%%%%%%%%%%%%%%%%%%%%%%%%
\section{Introduction}
\label{Sec:Intro}

Extinction in astronomy describes how much light is absorbed (or
scattered) by dust grains in the medium between the source and the
observer. Locally, extinction is measured by comparing the light
emitted from stars of similar spectral type, for which differential
variations from interstellar dust can be assessed. Local measurements
in the Milky Way reveal a general trend in the extinction law although
with significant scatter \citep[see, e.g.][]{FM:90}. The behaviour of
the extinction law reveals the overall properties of the dust grains
in the interstellar medium \citep{Draine:11}. Closer to home, light
scattered by the small air molecules in the atmosphere follow a
typical $\propto\lambda^{-4}$ law that explains the blueness of the
day-time sky. The typical Milky Way extinction law implies a shallower
scaling dependence within the optical and near-infrared window,
$\propto\lambda^{-1.5}$\citep{CCM:89,Fitz:99}. This effective law
results from the integration of the contribution from all grains in
the dust component. A distribution with smaller grain sizes would make
this law steeper, and a distribution with larger grains would induce a
shallower dependence with wavelength, approaching the Mie
regime \citep{Draine:11}, where there is no wavelength dependence
(i.e. clouds are white). One classical parameterisation of this smooth
wavelength dependence is the ratio of total-to-selective extinction,
$R_V\equiv A_V/E(B-V)$, where $A_V$ is the extinction, measured in
magnitudes, in the $V$ band, and $E(B-V)=A_B-A_V$ is the colour
excess, i.e. the additional reddening, measured as a $B-V$ colour,
caused by dust. In addition to this smooth trend in the extinction
law, there are some features characteristic of a specific
component. In our analysis -- extended to the near UV and optical
spectral window -- the most remarkable feature is the NUV bump at
2,175\AA\ \citep{Stecher:69}, see Fig.~\ref{fig:atten}. Although the
component responsible for this feature is not confirmed, it is
possibly carbonaceous material, graphite, or PAH
molecules \citep{StecherDonn:65}, although a more complex mixture is
possible \citep{Brad:05}.  NUV observations of the Milky Way
extinction curve consistently find this feature
\citep[see, e.g.][]{FM:86}. 
In contrast, both lower metallicity regions, such as the Small
Magellanic Cloud \citep{Pei:92}, and starbursting
galaxies \citep{Calz:01}, lack the NUV bump.

An extinction law is a function that takes into account the
wavelength-dependent obscuration due to absorption and scattering
processes as light traverses a distribution of dust particles. One can
therefore think of it as a foreground screen between the source and
the observer. Such a scenario is a valid representation when observing
stars in the Milky Way. However, when we extend the analysis of dust
to distant, mostly unresolved, galaxies, this process is complicated
by the distribution of the dust, the variation of the composition
within the observed region, and the underlying stellar populations. It
is common to refer to this integrated effect of dust in galaxies as
attenuation. Galaxies with the same dust properties, hence with the
same extinction, can yield different attenuation laws \citep[see,
e.g.][]{Granato:00}. For instance, the absence of the NUV bump in the
attenuation law of starbursting galaxies
\citep{Calz:01} can either be caused by an intrinsic lack of the
component that causes this bump, or by the superposition of several
stellar populations with different dust
geometries \citep{Panuzzo:07}. Although it is beyond the scope of this
paper to disentangle extinction and attenuation, one should bear in
mind this subtlety in the interpretation of the observational
constraints.

This work focuses on the dust component in the nearby starburst galaxy
M82 (NGC\,3034) located at a distance of $\sim$3.3-3.5\,Mpc 
\citep{Dalcan:09,Foley:14}, with a dynamical mass
of $\approx\!10^{10}$M$_\odot$ \citep{Greco:12}. The ongoing strong
star formation in this galaxy is believed to have been triggered by a
close encounter with the more massive galaxy M81, located in the same
group. UV light extends out from the central regions of the galaxy
where star formation is the strongest, revealing the presence of dust
entrained in the supernov\ae-driven wind. 
In \citet[][hereafter HF14]{Hutton:14} we found that the dust obscuration in the
NUV required the presence of an NUV bump, with the puzzling result
that the prototypical starburst attenuation law of \citet{Calz:01} ---
which features no NUV bump --- gave poorer fits than the standard Milky
Way extinction of \citet{Fitz:99}. This paper extends this issue by
probing in more detail the dust properties of the interstellar medium
(ISM) of M82, as a function of galactocentric distance.

Literally, as the accepted HF14 paper went to press, an enthusiastic
and diligently-led team of students at UCL's University of London
observatory discovered SN2014J, a type~Ia supernova, within 1\,kpc of
the central starburst of M82 \citep{SN2014J}. A number of papers
followed the evolution of the supernova
\citep[see, e.g.,][]{Goo:14,Zheng:14}, and, most relevant to this
paper, the properties of the dust 
\citep[see, e.g.][]{Aman:14,Foley:14,Kawabata:14,Patat:14}. 
We probe here the dust properties in the ISM of M82
pre-SN2014J, an issue of relevance to the understanding of the  
progenitor system of this supernova event.

%%%%%%%%%%%%%%%%%%%%%%%%%%%%%%%%%%%%%%
%%%%%%%%%%   TABLE 1   %%%%%%%%%%%%%%%
%%%%%%%%%%%%%%%%%%%%%%%%%%%%%%%%%%%%%%
\begin{table*}
\caption{Parameter range of the grid of simple stellar population
models used to derive the effective dust-related properties.}
\label{tab:params}
\begin{tabular}{lccr}
\hline
\multicolumn{4}{c}{Single Burst (1SSP): $\psi(t)\propto\delta(t-t_0)$}\\
\hline
Observable & Parameter & Range & Steps\\
\hline
Age & log(t$_0$/Gyr) & $[-2,+0.9]$ & 32\\
Metallicity & $\log Z/Z_\odot$ & $[-1.5,+0.3]$ & 8\\
Total to selective extinction ratio & $R_V$ & $[0.5,4.5]$ & 32\\
NUV Bump strength & B& $[0,1.3]$ & 32 \\ 
Colour excess & E(B-V) & $[0,1.5]$ & 32 \\ 
\hline
\multicolumn{3}{r}{Number of models} & 8,388,608\\
\end{tabular}
\end{table*}
%%%%%%%%%%%%%%%%%%%%%%%%%%%%%%%%%%%%%%%%%%%%%%%%

The structure of the paper is as follows: the NUV-to-Optical data used
for the analysis are presented in \S\ref{Sec:Data}, followed by the
modelling of the dust attenuation law in \S\ref{Sec:DustLaw}. The
methodology followed to constrain the dust parameters via stellar
population synthesis is explained in \S\ref{Sec:method}, with the
results shown in \S\ref{Sec:M82Dust}, along with comparisons with the
recent constraints on the dust properties from
SN2014J. Finally, \S\ref{Sec:Conc} closes with the conclusions of our
analysis.

%%%%%%%%%%%%%%%%    Figure 2   %%%%%%%%%%%%%%%%%
%%%%%%%%%%%%%%%%%%%%%%%%%%%%%%%%%%%%%%%%%%%%%%%%
\begin{figure*}
\begin{center}
\includegraphics[width=8.7cm]{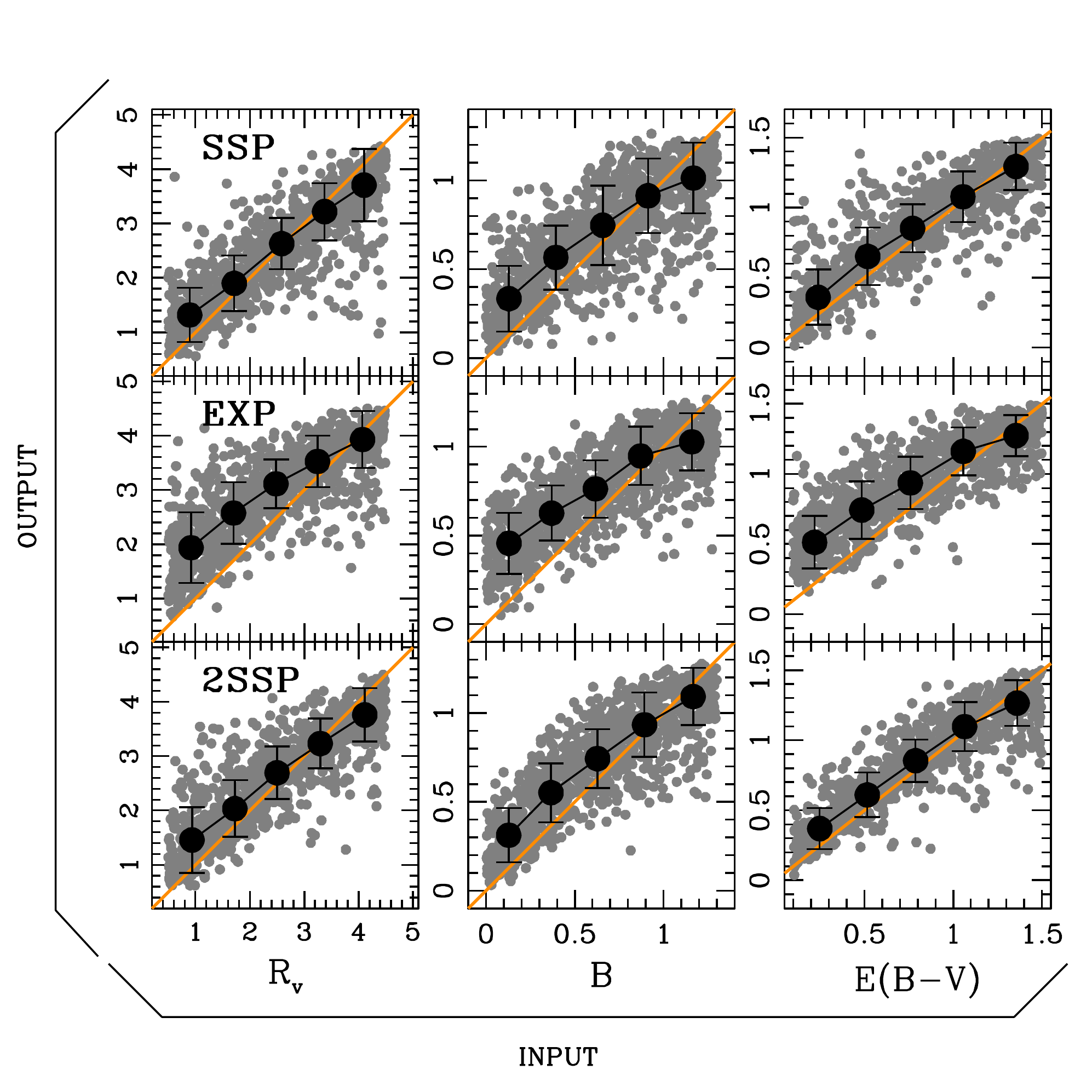}
\includegraphics[width=8.7cm]{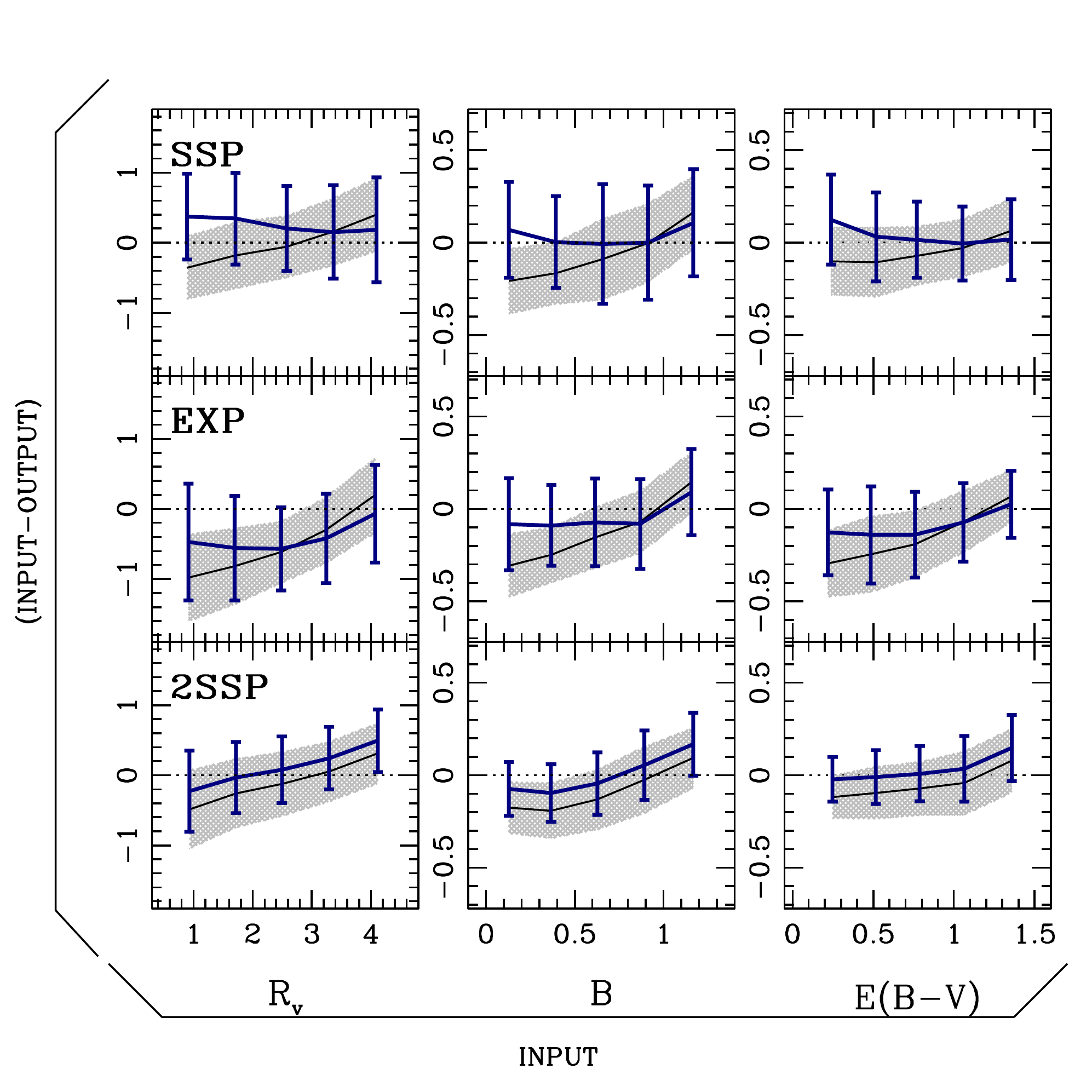}\\
\end{center}
\caption{Results from simulated data. The 
same method used for the analysis of the NUV+optical photometry of M82
is applied to a set of simulated spectra. {\sl Left:} Direct
comparison between input (i.e. true) and output (i.e. retrieved)
parameters. Each row represents a set of 1,000 simulations where the
adopted star formation history is a simple stellar population (SSP,
top); an exponentially decaying star formation rate (EXP, middle) or a
two-burst history (2SSP, bottom). The grey dots are individual
results, whereas the big circles and error bars are median and RMS
values, binned at fixed number of data points per bin. 
{\sl Right:} The comparison is given with respect to the difference
between retrieved and true values, expressed as a fraction of the uncertainty.
The grey shaded regions and lines are the raw results, whereas the blue error 
bars correspond to a correction based on a calibration of the output using these
simulations (see Tab.~\ref{tab:corr} and text for details).}
\label{fig:sims}
\end{figure*}
%%%%%%%%%%%%%%%%%%%%%%%%%%%%%%%%%%%%%%%%%%%%%%%%

%%%%%%%%%%%%%%%%%%%%%%%%%%%%%%%%%%%%%%%%%%%%%%%%
\section{Data}
\label{Sec:Data}

We use the same dataset as in HF14. In a nutshell, deep NUV images
were retrieved from the HEASARC archive of {\sl Swift}/UVOT.  The
Ultraviolet/Optical Telescope \citep[UVOT,][]{Roming:05} is a 30\,cm
instrument working as a photon counter, with three passbands in the
NUV region, straddling the position of the 2,175\AA\ bump. We refer
the reader to \citet{Poole:08} and \citet{Breeveld:10} for details of
the properties, mode of operation and data reduction procedure of UVOT
imaging. We retrieved imaging data of M82 with total exposure times of
10.2, 13.9 and 8.7\,ks in the UVW2, UVM2, and UVW1 passbands,
respectively. The data were taken between 2008 and 2012, therefore our
photometry corresponds to the galaxy before the SN2014J event. In
addition to the UVOT data, we retrieved optical images from the SDSS
DR8 server \citep{sdss:dr8}. All images were registered to a common
grid with a 0.5\,arcsec pixel size (i.e. the UVOT reference). The
optical (SDSS) images were convolved to a common spatial resolution
with a Moffat profile. We note that the effective resolution of the
UVOT images is $\sim\!2.5-3$\,arcsec.  Photometry is performed within
5\,arcsec (radius) apertures, following the standard calibration of
the UVOT camera
\citep{Breeveld:10}. The apertures were chosen to follow the
disc of M82 (see fig.~2 of HF14). We also retrieved the GALEX/$FUV$
images of M82 from the MAST archive, creating new sets of photometric
measurements by convolving the UVOT data to the (poorer) resolution of
GALEX\footnote{Those lower resolution images were only used for the
$FUV-UVW2$ colour} ($\sim\!5$\,arcsec).  Therefore, our
combined dataset comprises photometry in the
$FUV,UVW2,UVM2,UVW1,u,g,r,i,z$ passbands.  We refer the interested
reader to HF14 for further details about the dataset.

There are, however, two differences with respect to the HF14
data. Here, we make use of an updated estimate of the foreground
(Milky Way) reddening towards M82. In HF14 we took the value derived
by \citet{Schlafly:11}, as given in the NED database, $E(B-V)\!=\!0.140$\,mag.
Following the suggestion of \citet{Dalcan:09}, updated by \citet{Foley:14},
we choose a lower estimate of $0.054$\,mag. This correction is due to the fact
that M82 affects the reading of the large scale dust maps used for the
determination of foreground reddening. This change affects the correction
made to the observed photometry, and the derivation of the dust parameters,
since in this paper, the amount of correction from the Milky Way 
-- which follows a $R_V\!=\!3.1$ \citet{Fitz:99} law -- is significantly
reduced. Nevertheless, we find that our results are compatible with
HF14.  In addition, we use 3.3\,Mpc for the distance to M82, for a
more consistent comparison with recent
papers \citep[e.g.][]{Foley:14}. This change only affects the
translation from angular distances to projected physical distances. The
photometric apertures adopted in this paper map into a projected 160\,pc diameter.

%%%%%%%%%%%%     Figure 3      %%%%%%%%%%%%%%%%%
%%%%%%%%%%%%%%%%%%%%%%%%%%%%%%%%%%%%%%%%%%%%%%%%
\begin{figure*}
\begin{center}
\includegraphics[width=5.8cm]{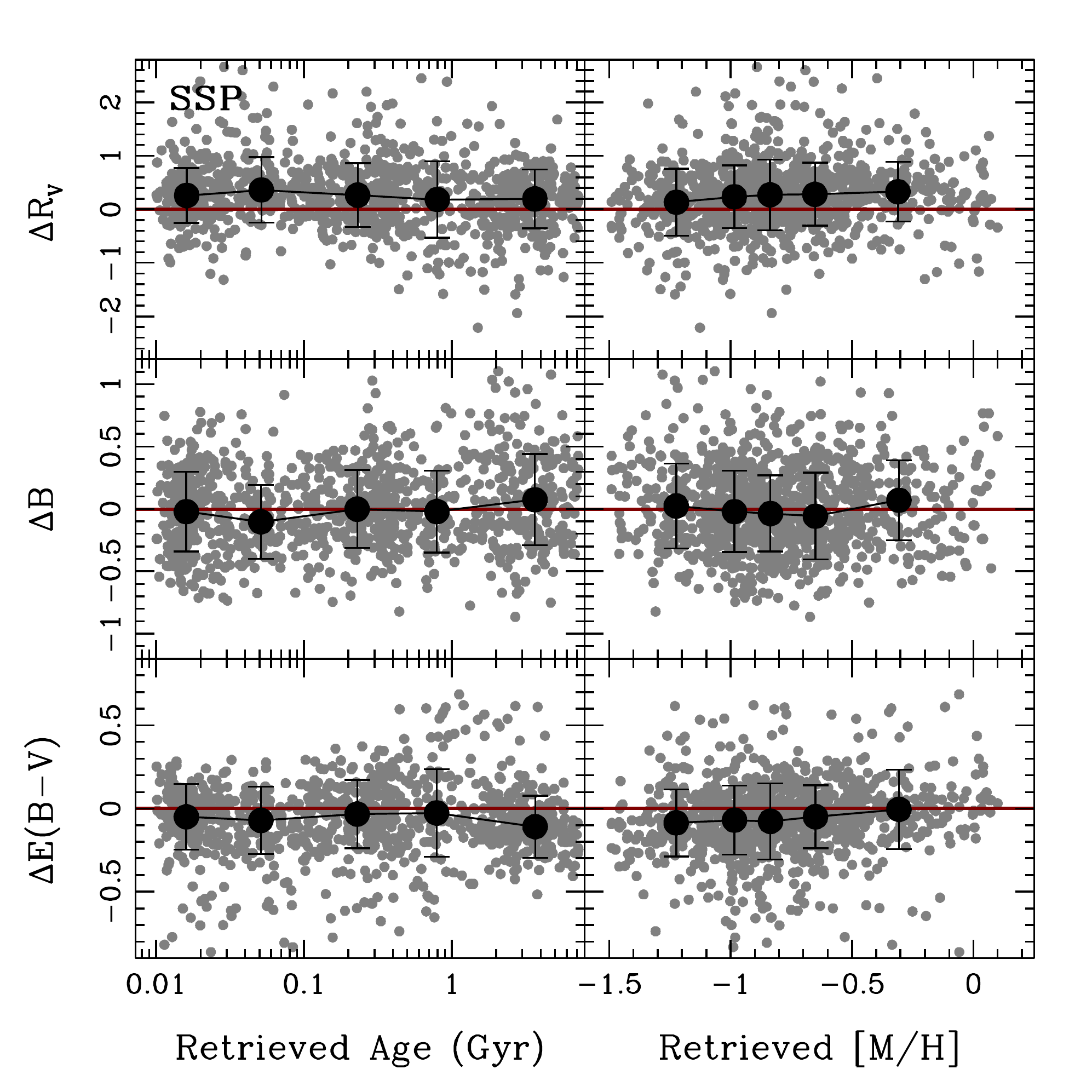}
\includegraphics[width=5.8cm]{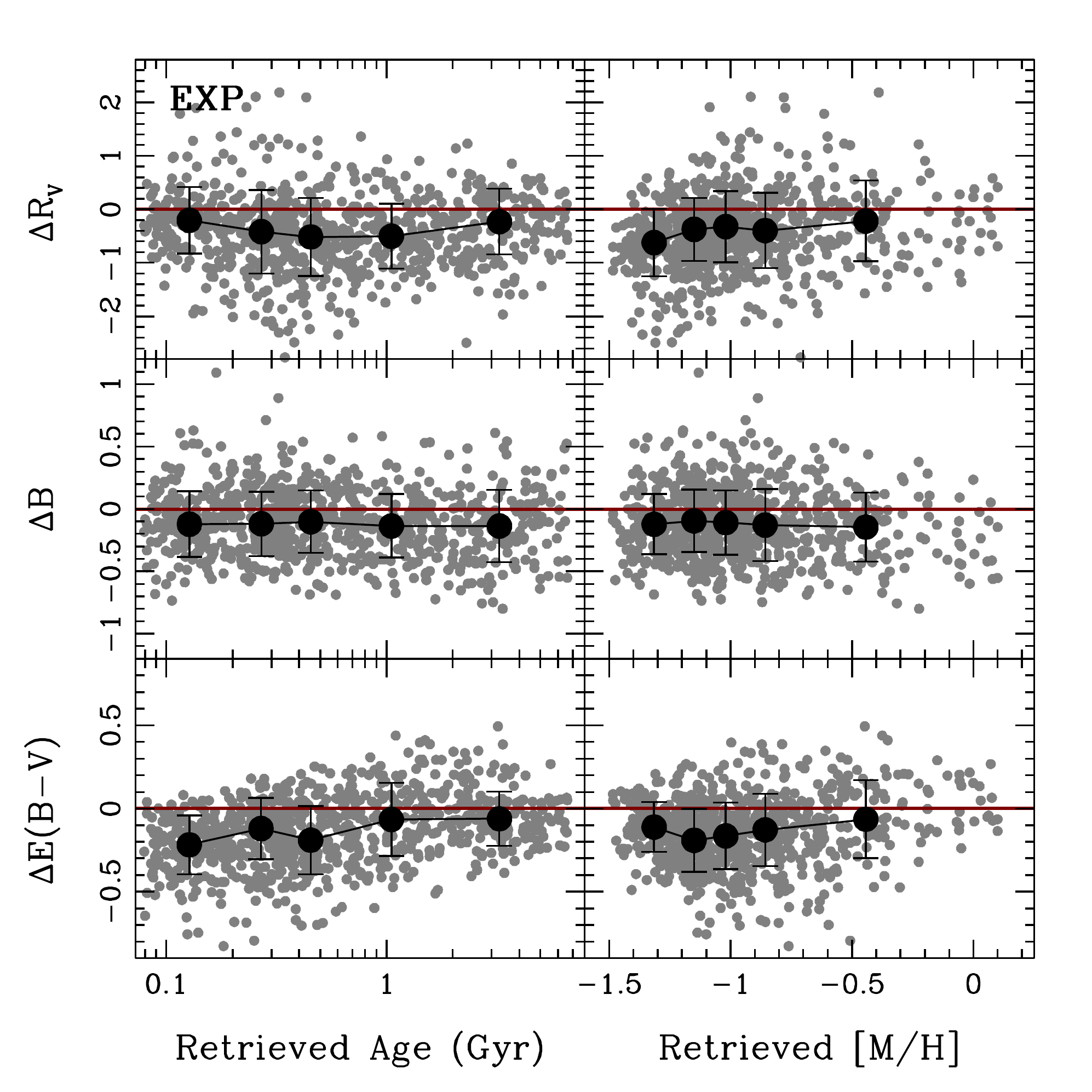}
\includegraphics[width=5.8cm]{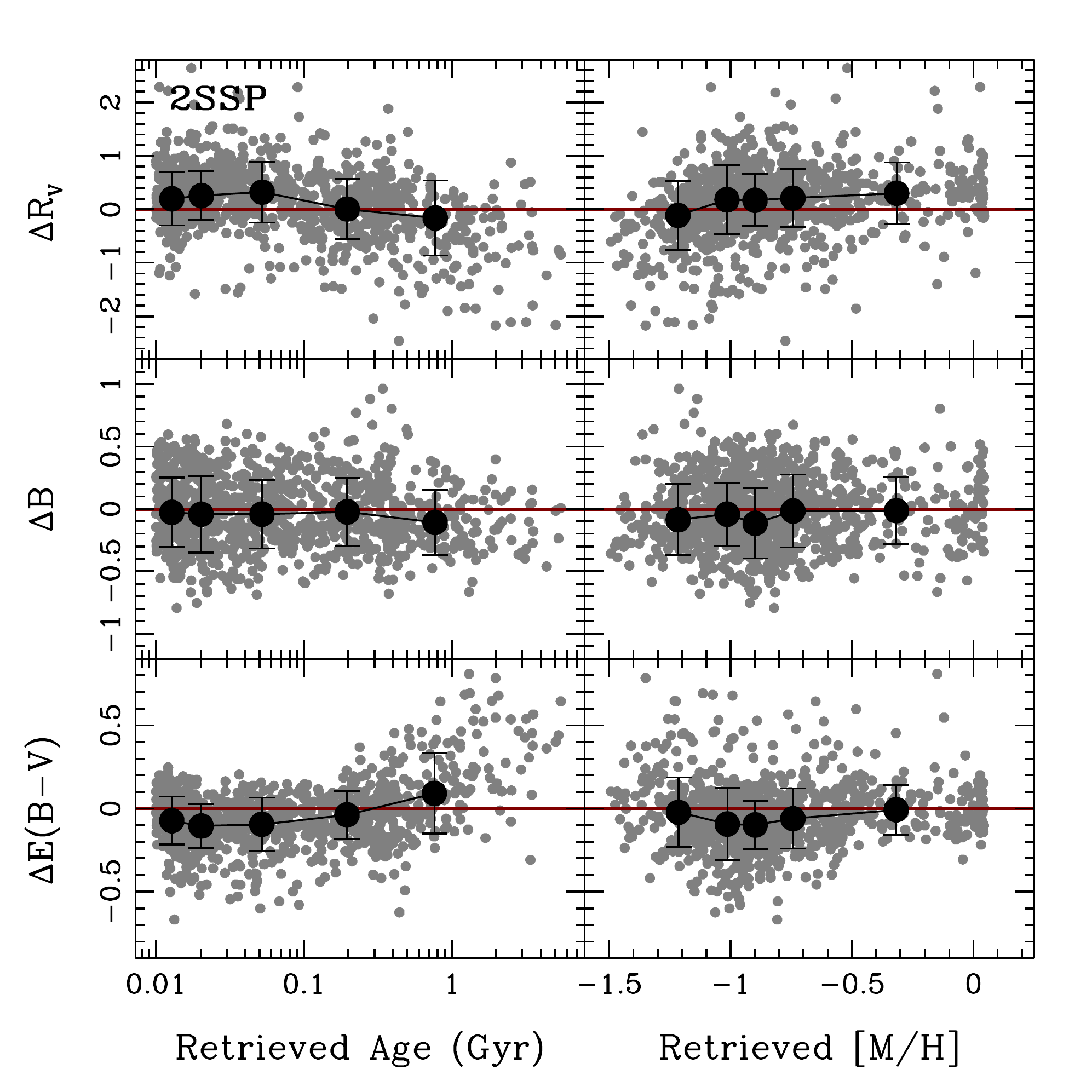}\\
\end{center}
\caption{Trend of the retrieved dust parameters with 
the age and metallicity derived from the same likelihood
function. The offset is defined as $\Delta\pi\equiv \pi({\rm
in})- \pi({\rm out})$, where $\pi$ represents any of the three
dust-related parameters. There is no significant bias in the dust
parameters with respect to the underlying stellar populations.}
\label{fig:simstZ}
\end{figure*}
%%%%%%%%%%%%%%%%%%%%%%%%%%%%%%%%%%%%%%%%%%%%%%%%

%%%%%%%%%%%%%%%%%%%%%%%%%%%%%%%%%%%%%%%%%%%%%%%%
\section{Modelling the attenuation law}
\label{Sec:DustLaw}

We follow the parameterisation of the attenuation law of 
\citet[][hereafter CSB10]{CSB:10}. This function depends on
two basic parameters: $R_V$ (defined in \S\ref{Sec:Intro}), and $B$,
which controls the strength of the NUV bump.  Fig.~\ref{fig:atten}
shows a few realizations of the CSB10 function, along with the
standard Milky Way attenuation law from \citet{Fitz:99}, assuming
$R_V=3.1$.  Notice that within the NUV spectral window (left-hand
panels), the UVOT filters sample the bump, with UVM2 located in the
middle of the feature and UVW2/UVW1 straddling it on the blue and red
sides, respectively. In addition, the GALEX/FUV filter includes an
additional measurement blueward of the bump.  The case $B$=1 corresponds
very nearly to the standard Milky Way extinction curve, by
construction. We follow a simple one-zone model for the dust
component, assuming the light from the stellar populations is affected
by a foreground dust screen. Other methods \citep[e.g.,][]{Granato:00}
would result in a more complex set of parameters, whereas we focus
here on an ``effective'' attenuation law integrated to all the light
within the aperture \citep[see also][]{Burg:05}.

In HF14, we compared the NUV and optical photometry of M82 with
a range of stellar population models, and two typical
attenuation curves, the above mentioned Milky Way standard of
\citet{Fitz:99}, and the law of \citet{Calz:01} usually 
applied to star-bursting systems such as M82. We found that the latter
-- which lacks the presence of an NUV bump -- gave poorer fits to the
NUV photometry with respect to a Milky-Way law. In this paper, we
explore in more detail this issue by use of the parameterization of
CSB10.

In addition to the NUV bump, the photometric data can be used to
constrain the parameter $R_V$, which depends on the composition of the
dust grains. A standard Milky Way law assumes $R_V=3.1$, although with
significant scatter, roughly between 2.2 and
5.8 \citep{Fitz:99}. Lower (higher) values of $R_V$ will require a
distribution of smaller (larger) dust grains, with respect to the
Milky Way standard. In HF14, we studied the extraplanar light
originated from the reflection of the stellar light in the galaxy,
that impinged on the dust grains above and below the galactic plane.
These dust grains were entrained in the material ejected by the strong
galactic winds triggered by the starburst. We fitted the photometric
data assuming the absorption/scattering processes are described by a
power law ($\propto\lambda^{-x}$), finding a value ($x\!=\!1.53\pm
0.17$). Note, however, that this case pertains to the scattering of
light from dust grains in the extraplanar wind region, as they are
illuminated by the galaxy.  In contrast, we focus in this paper on the
efective attenuation law in the main body of the galaxy.

The right-hand panels of Fig.~\ref{fig:atten} show a wide NUV/optical
spectral window, focusing on variations of $R_V$. Note how higher
values of $R_V$ -- as expected in an ISM with larger dust grains --
result in a flatter attenuation law, closer to the Mie
regime. Therefore, the use of a wide spectral range (NUV and optical
photometry) will allow us to constrain this parameter.

%%%%%%%%%%%%%%%%   Figure 4  %%%%%%%%%%%%%%%%%%%
%%%%%%%%%%%%%%%%%%%%%%%%%%%%%%%%%%%%%%%%%%%%%%%%
\begin{figure}
\begin{center}
\includegraphics[width=8.6cm]{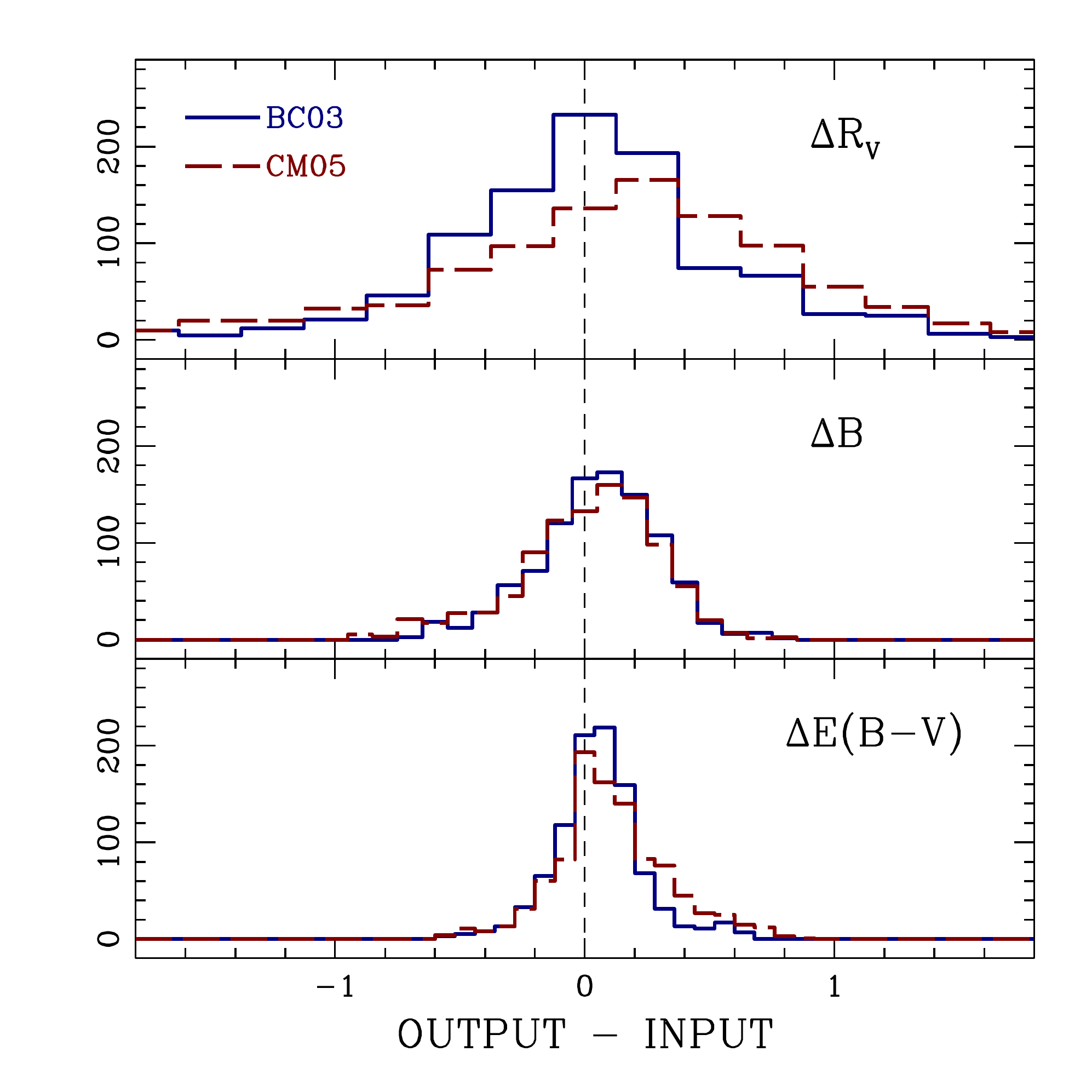}
\end{center}
\caption{Histogram comparing the difference between the ``true''
(input) and the retrieved (output) dust-related parameters, for two
choices of population synthesis models. The methodology is the same in
both cases, using a grid of SSP models from \citet{BC03}. The blue
solid histograms show the parameter retrieval accuracy when using a
range of simulations constructed from \citet{BC03} models. The red
dashed histograms correspond to simulations assembled from the models
of \citet{CM:05}. No significant bias is found (see text for
details).}
\label{fig:CM05_comp}
\end{figure}
%%%%%%%%%%%%%%%%%%%%%%%%%%%%%%%%%%%%%%%%%%%%%%%%

%%%%%%%%%%%%%%%%%%%%%%%%%%%%%%%%%%%%%%%%%%%%%%%%
\section{Constraining the Dust Values}
\label{Sec:method}

In order to constrain the dust attenuation parameters, we need to
quantify the effect of the other, ``nuisance'', parameters related to
the underlying stellar population. Our methodology involves a
comparison of the observed NUV and optical photometry with a large
grid of synthetic populations from the models
of \citet{BC03}. Tab.~\ref{tab:params} shows the grid of 
8,388,608 models used in this paper. They correspond to simple
stellar populations (SSP, representing a population with a single age
and metallicity) for a standard initial mass
function \citep{Chab:03}. Within our wavelength interval, dust simply
introduces a smooth, long (spectral) range behaviour (aside from the
NUV bump). Therefore it is possible to factor out the attenuation law
from the different trends introduced in the photometry by the age and
metallicity distribution of the populations. To prove this point, we
ran three sets of simulations, each one with a different assumption
about the star formation history: i) SSP (as above); ii) EXP: a
composite of stellar populations following an exponentially decaying
star formation rate, choosing the formation epoch, decay timescale and
metallicity as free parameters of the populations; iii) 2SSP, a
superposition of two SSPs, where the age of the young and old
components, the fractional contribution, by mass, of the young
component, and the metallicity are free parameters. These three
choices encompass a wide range of both smooth and bursty star
formation histories. This exercise serves to understand whether
systematic biases are being introduced because of a methodology where
the analysis is based on a grid of SSPs.

The derivation of the model parameters follows a Bayesian
approach, using a $\chi^2$-based likelihood as a probability
distribution function (${\cal L}\propto e^{-\Delta\chi^2/2}$) from
which we derive the errors as the 68\% confidence level of this
distribution. $\Delta\chi^2$ is defined as $\chi^2-\chi^2_{\rm min}$,
and $\chi^2$ is given by:
\begin{equation*}
\chi^2(\pi_i)\equiv\sum_j\frac{\Large[ m_j^{\rm OBS}-m_j^{\rm MODEL}(\pi_i)\Large]^2}{\sigma^2(m_j)},
\end{equation*}
where $\{\pi_i\}$ represents all possible choices of the five
parameters listed in Tab.~\ref{tab:params}. $m_j^{\rm OBS}$ and
$m_j^{\rm MODEL}$ are the observed and model magnitudes, respectively,
as follows: \{FUV, UVW2, UVM2, UVW1, $u$, $g$, $r$, $i$,
$z$\}. Finally, $\sigma(m_j)$ is the observational errors for each
magnitude.  In addition to the parameters listed in
Tab.~\ref{tab:params}, we include a ``normalization'' parameter that
assumes the model SDSS $g$ band magnitude lies somewhere between
$g_{\rm OBS}\pm 0.1$\,mag; where $g_{\rm OBS}$ is the observed
magnitude in the $g$ band for each aperture. The extracted parameters
are then computed using the likelihood as a probability distribution
function, namely:
\begin{equation}
\langle\pi_i\rangle = \frac{\int d\pi_1\cdots d\pi_5{\cal L}(\pi)\pi_i}
{\int d\pi_1\cdots d\pi_5{\cal L}(\pi)}
\end{equation}
The priors applied in the analysis are simply the range of parameters
explored, as shown in Tab.~\ref{tab:params} (i.e. one can consider
them as ``flat priors'', giving equal probability within the range
stated).

%%%%%%%%%%%%      Figure 5     %%%%%%%%%%%%%%%%%
%%%%%%%%%%%%%%%%%%%%%%%%%%%%%%%%%%%%%%%%%%%%%%%%
\begin{figure*}
\begin{center}
\includegraphics[width=7.4cm]{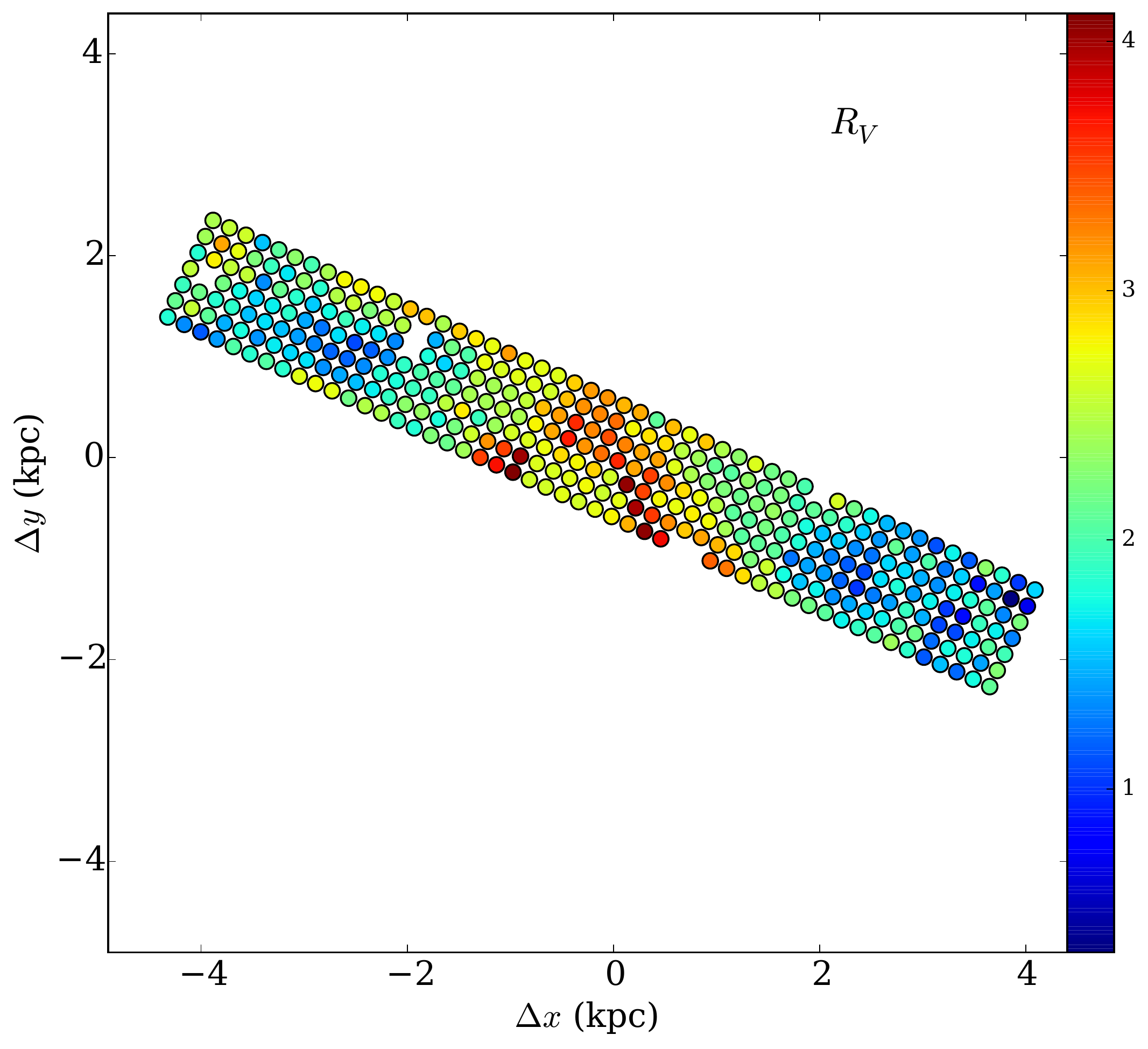}
\includegraphics[width=7.5cm]{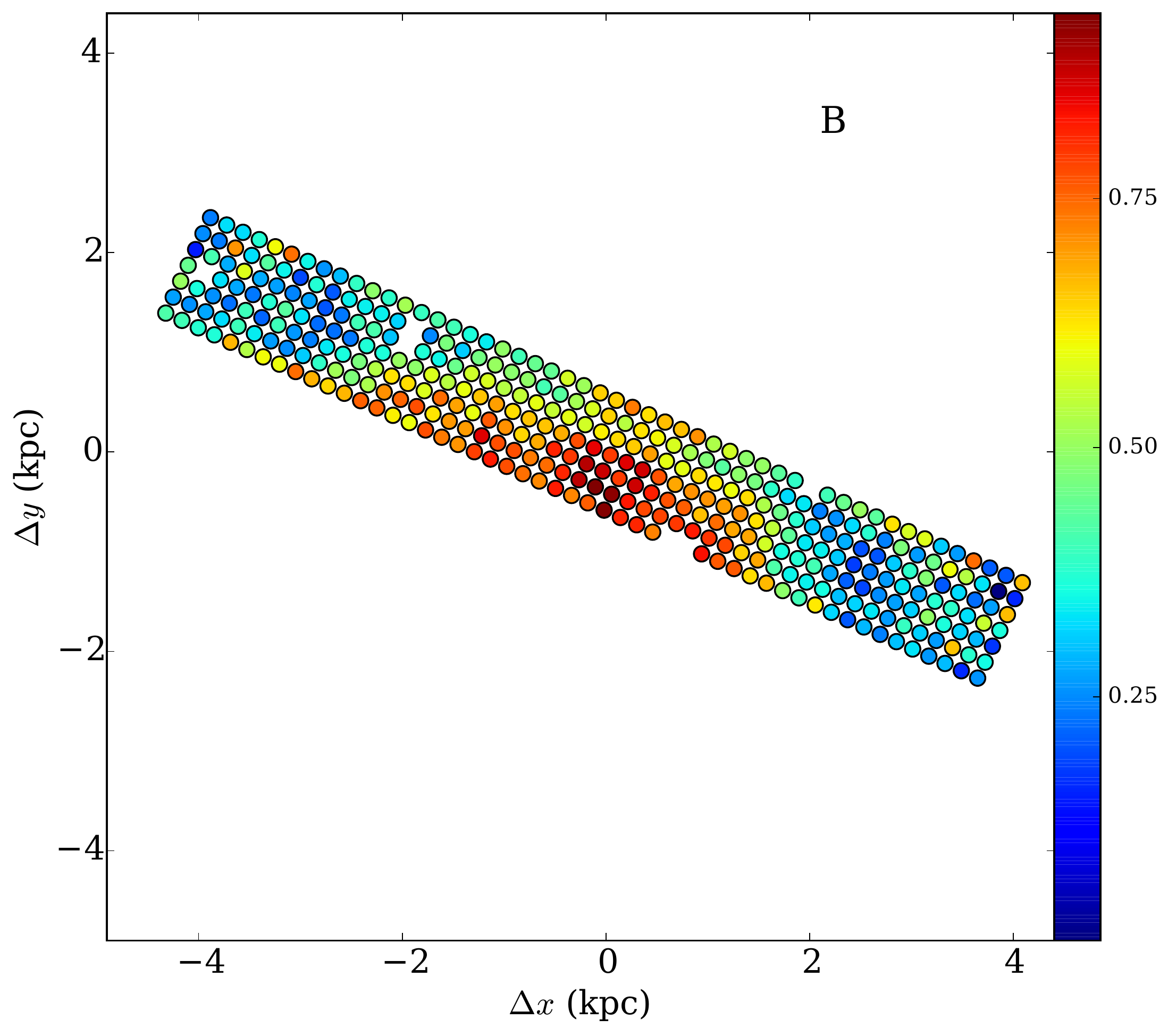}\\
\includegraphics[width=7.5cm]{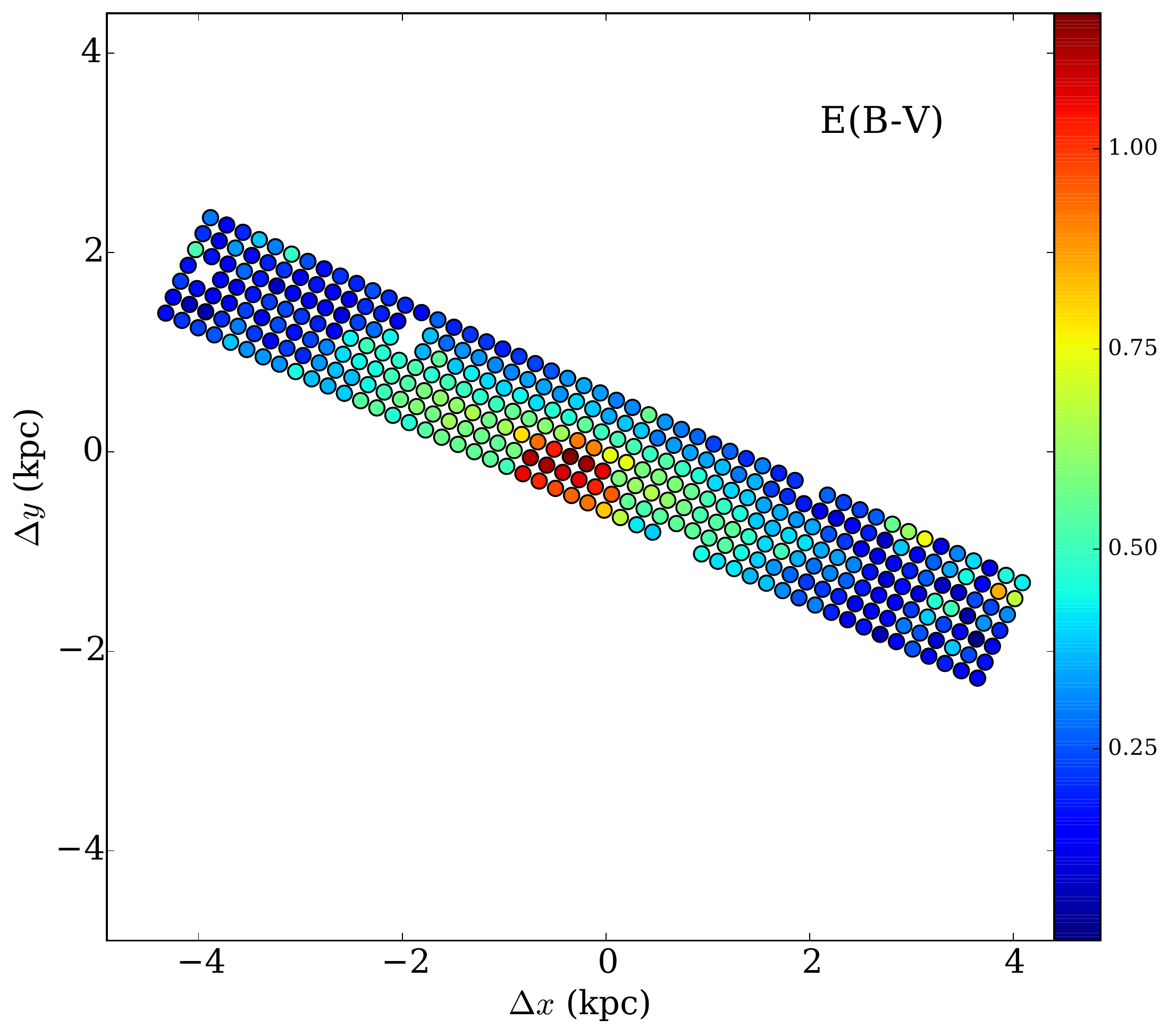}
\includegraphics[width=7.5cm,height=6.7cm]{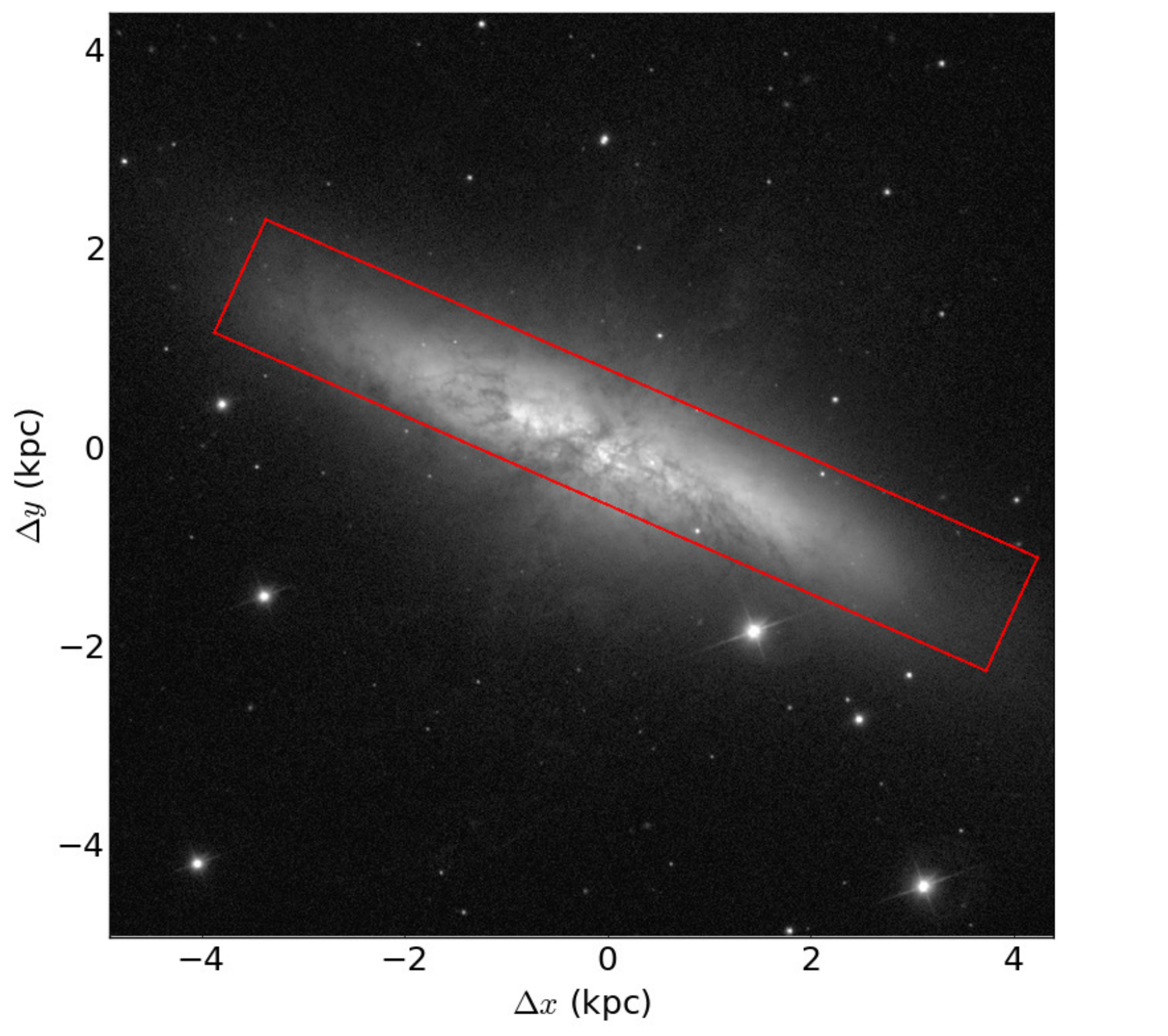}
\end{center}
\caption{Two-dimensional distribution of the retrieved 
values of the parameters that define the dust attenuation law used in
this paper (see text for details). The ratio of total-to-selective attenuation
 ({\sl top-left}), NUV bump strength ({\sl top-right}) and colour
excess ({\sl bottom-left}) are shown as a colour-coded map, with the
contiguous colour bars showing the mapping in each panel. The x-y coordinates are
given as physical projected distances in kpc. The bottom-right panel
shows a $g$-band image from SDSS-DR10 at the same scale.}
\label{fig:maps}
\end{figure*}
%%%%%%%%%%%%%%%%%%%%%%%%%%%%%%%%%%%%%%%%%%%%%%%%

Fig.~\ref{fig:sims} shows the results for 3$\times$1,000 simulations
corresponding to each of the three formation histories. The simulated
data explore a wide range of ages, metallicities, and dust properties,
comparable to those found in M82 (see fig.~5 of HF14). Our analysis
marginalizes over the stellar population properties, and produces the
extinction-related constraints on -- from left to right -- $R_V$, $B$
and $E(B-V)$. The first two parameters are intrinsic to the effective
dust attenuation law, whereas the latter can be considered as a
normalization factor, reflecting the amount of dust present. The
panels on the left compare input value (abscissa) versus retrieved
value (ordinate). An orange line in each panel gives the 1:1
correspondence. The grey dots are results from individual estimates,
whereas the black dots and error bars mark the median and root-mean
square values, respectively, when binned at fixed number of data
points per bin. Note that no large bias is obtained, regardless of the
use of different star formation histories (top to bottom).  The
derivation of the NUV bump strength only uses the UV photometry for
the definition of the likelihood, since the optical photometry is
fully insensitive to the presence of this bump (Fig.~\ref{fig:atten}
right-hand panel), and would therefore ``wash-out'' the constraining
power in the comparison. 

We use the comparison with the simulated data to derive a mean linear
correction between input and output parameters. The results of the
fits are shown in Tab.~\ref{tab:corr}. Since one cannot determine the
best star formation history corresponding to the real data, we decided
to choose for the general correction the one that gave the best
compromise among the three sets of star formation histories, i.e. we
used the results from all three sets of simulations for the fit. The
right-hand panels of Fig.~\ref{fig:sims} illustrate the validity of
this correction and the retrieved error bars. They compare the
difference between the ``true'' and the extracted value
(``input''$-$``output'' in the figure). This comparison allows us to
determine whether the quoted error bars are valid. The shaded regions
correspond to the original data, whereas the blue lines with error
bars give the results for the corrected parameters. Note that the
corrections are relatively minor with respect to the error bar, and no
significant bias is apparent when the corrections are taken into
account. Nevertheless, we will show in the next figures our results
with and without this correction. The comparison with simulated data
therefore confirms that it is possible to factor out the attenuation
law from the properties of the underlying, unresolved stellar
populations, by use of a large library of simple stellar populations.

%%%%%%%%%%%%%%%%   Figure 6  %%%%%%%%%%%%%%%%%%%
%%%%%%%%%%%%%%%%%%%%%%%%%%%%%%%%%%%%%%%%%%%%%%%%
\begin{figure}
\begin{center}
\includegraphics[width=8.5cm]{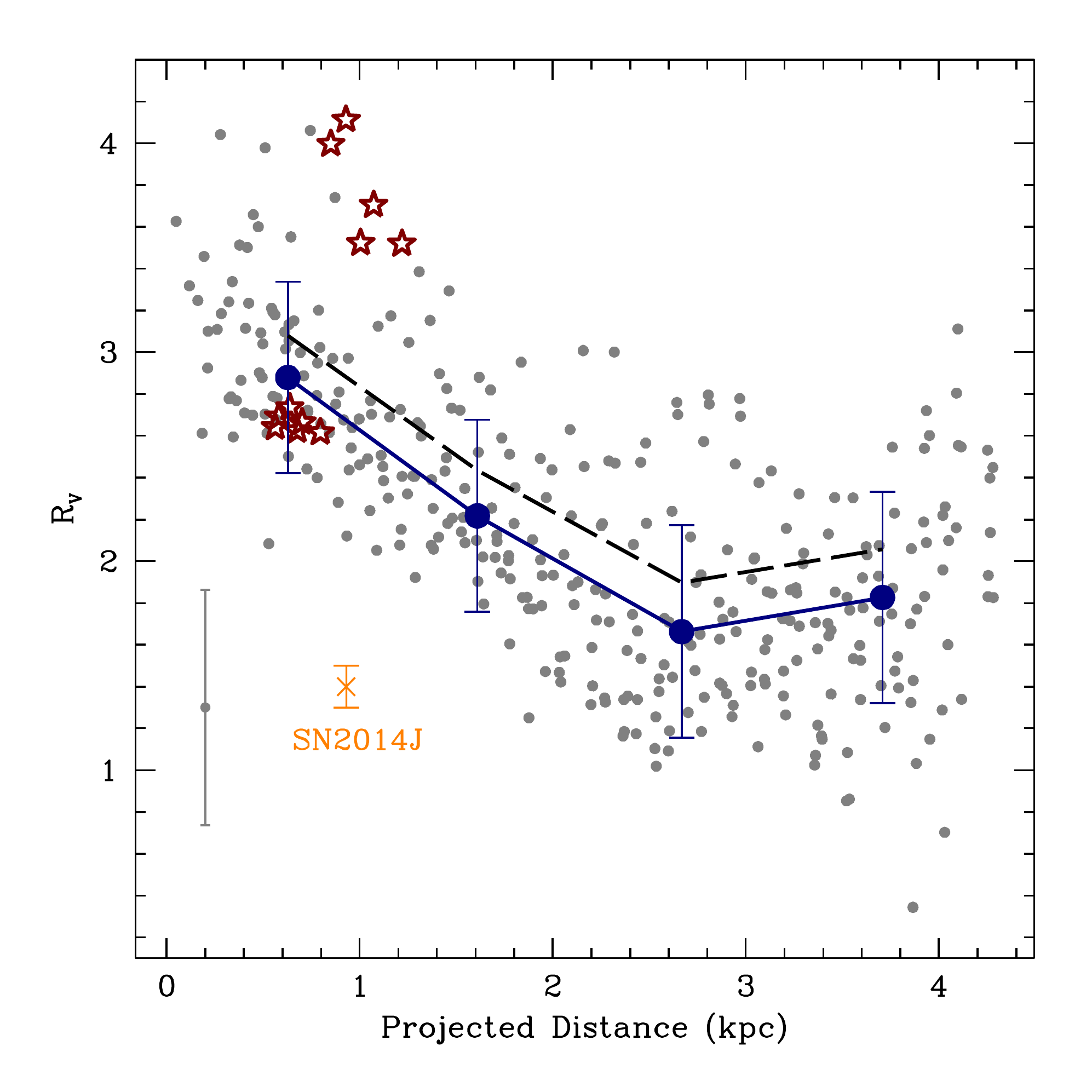}
\end{center}
\caption{Retrieved values for the ratio of total-to-selective
extinction $R_V$ as a function of projected galactocentric
distance. The grey dots represent individual measurements, with a
characteristic error bar shown on the left-hand side of the
figure. The red stars are measurements within 30\,arcsec (i.e. 480\,pc)
of the position of SN2014J. The blue dots and error bars are the
median and root mean square values of the data, binned at a fixed
number of data points per bin. The black line shows the same result
if the retrieved parameters are not calibrated with respect to
the simulations (see Tab.~\ref{tab:corr}). The orange cross and error bar
correspond to the measurement in the region of SN2014J after the
event \citep{Aman:14} }
\label{fig:M82_Rv}
\end{figure}
%%%%%%%%%%%%%%%%%%%%%%%%%%%%%%%%%%%%%%%%%%%%%%%%

We also need to assess a possible bias between the dust
related parameters ($R_V$,$B$,$E(B-V)$) and the underlying stellar populations.
In Fig.~\ref{fig:simstZ} we compare the bias $\Delta\pi\equiv \pi({\rm in})-
\pi({\rm out})$ -- where $\pi$ represents any of the three dust-related
parameters -- as a function of the retrieved age and metallicity.
Note that the input parameters are, by construction, uncorrelated.
The figure shows that most of the comparisons give negligible
dependence on age and metallicity, or on the functional form of the
star formation history. This result is very important for the
interpretation of the radial gradients of the effective attenuation
law presented in the next section. We also note that the
values we retrieve for the SSP ages are consistent with 
\cite{Hutton:14}, where only two standard extinction laws
were considered. These results are also consistent with independent
estimates, giving ages around 0.4--1\,Gyr within the central 3\,kpc,
with older ages in the outer regions; and a colour excess 
$E(B-V)\sim 0.4$ with a significant increase in the central regions \citep[see,
e.g.][]{Mayya:06,RodMer:11}. As regards to metallicity, we emphasize
that broadband photometric analyses have large degeneracies with
respect to this parameter. Our simulations in fact show that of all
five parameters given in Tab.~\ref{tab:params}, metallicity is the only one that
fails at retrieving the input values, within a 0.5\,dex
uncertainty. However, we emphasize that such a result does not prevent
us from deriving unbiased estimates of the other parameters (see
Fig.~\ref{fig:sims}), and consider age and metallicity simply as
nuisance parameters in the derivation of the dust-related properties.
In addition, we have done tests, imposing the typical
metallicities expected in M82 \citep[e.g.][]{Origlia:04} as priors, and
the extracted values of $R_V$, $B$, and $E(B-V)$ are unaffected.
A partial version of the parameters derived from this analysis are
shown in Tab.~\ref{tab:results}. The complete version can be retrieved
online.

Finally, in order to assess the possible systematics caused by the
specific population synthesis model used, we create an additional set
of 1,000 simulations with the models of \citet{CM:05}. We note that
these models apply widely different prescriptions, including a
different treatment of the thermally-pulsing AGB phase, which can be
prominent in populations of ages $\sim 1$\,Gyr, therefore relevant for
the study of M82. We emphasize here that the aim in this paper is {\sl
not} to constrain the age or metallicity composition of the populations,
but to factor out the degeneracies caused by these parameters. In
order to show that our results are robust against different
prescriptions for the population synthesis models, we treat the
simulated data -- which use the \citet{CM:05} -- models, with the same
grid of SSPs -- which are based on the \citet{BC03} models.
Fig.~\ref{fig:CM05_comp} shows the histograms of the difference
between the ``true'' (i.e. the assumed value) parameter, and the
retrieved one, in both cases (BC03: for the simulated data based
on \citet{BC03} models; CM05 for the data based on the \citet{CM:05}
models). We conclude that our derivation of the dust-related
parameters is robust. 

%%%%%%%%%%%%%%%%   Figure 7  %%%%%%%%%%%%%%%%%%%
%%%%%%%%%%%%%%%%%%%%%%%%%%%%%%%%%%%%%%%%%%%%%%%%
\begin{figure}
\begin{center}
\includegraphics[width=8.5cm]{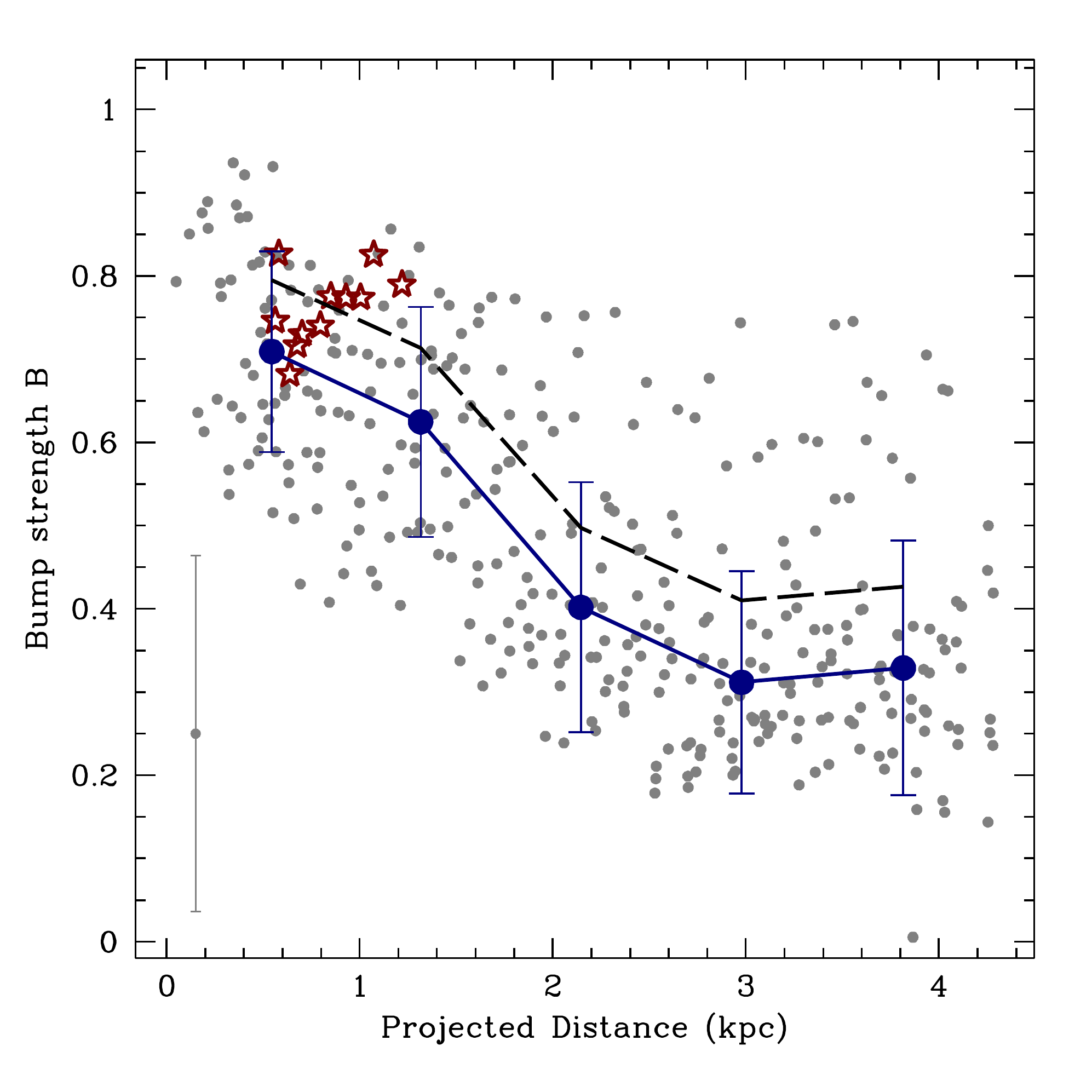}
\end{center}
\caption{The observational constraints on the NUV bump strength,
$B$, are plotted with respect to the projected galactocentric
distance.  The symbols are the same as in Fig.~\ref{fig:M82_Rv}.}
\label{fig:M82_B}
\end{figure}
%%%%%%%%%%%%%%%%%%%%%%%%%%%%%%%%%%%%%%%%%%%%%%%%

%%%%%%%%%%%%%%%%%%%%%%%%%%%%%%%%%%%%%%%%%%%%%%%%
\section{The attenuation law of M82}
\label{Sec:M82Dust}

\begin{figure}
\begin{center}
\includegraphics[width=8.5cm]{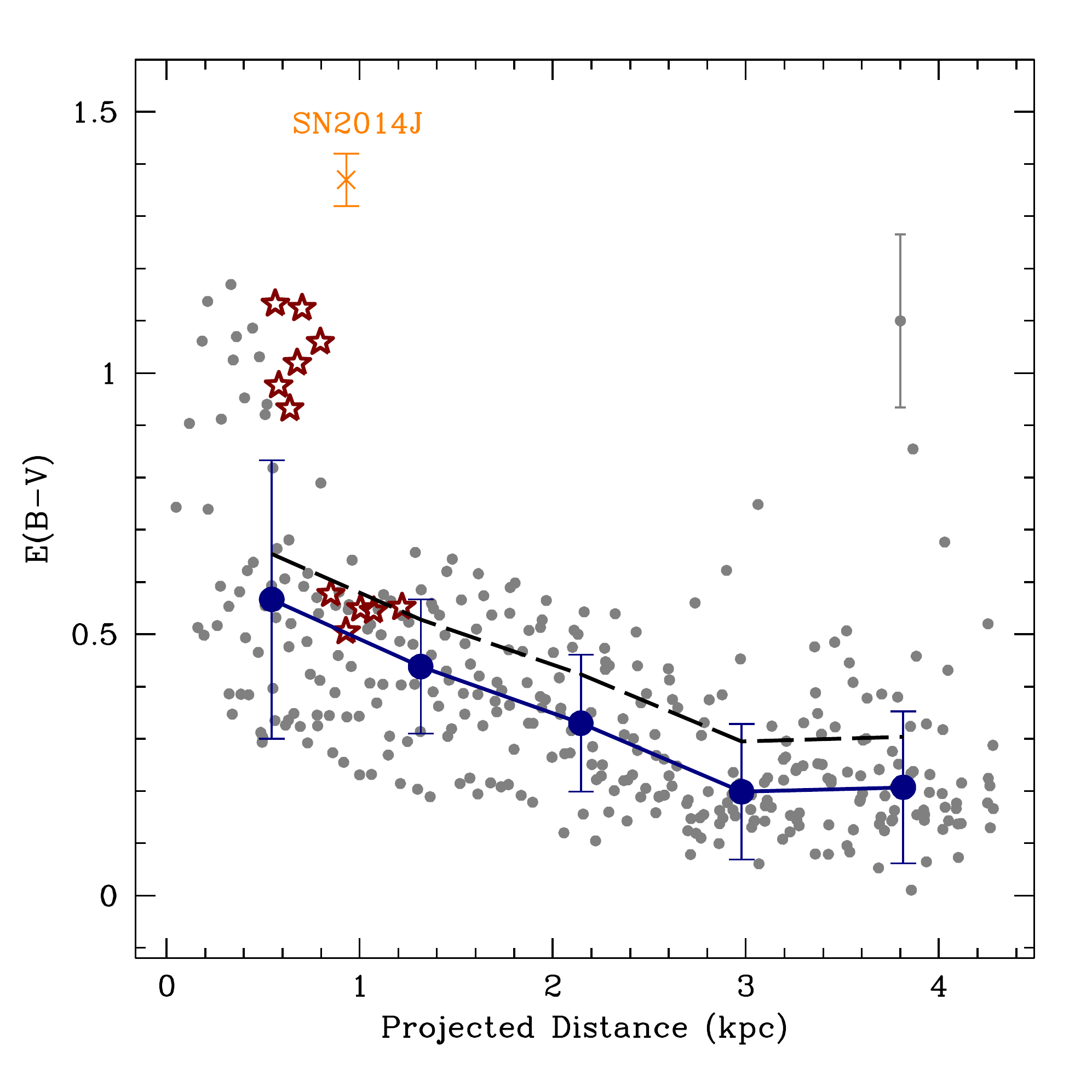}
\end{center}
\caption{The best fits to the colour excess, $E(B-V)$, are
shown against galactocentric distance. The symbols are the same as in
Fig.~\ref{fig:M82_Rv}.} 
\label{fig:M82_Ebv}
\end{figure}

Fig.~\ref{fig:maps} shows the spatial distribution of the
dust-related parameters constrained with the photometry. The results
are shown as colour-maps. The missing apertures in the distribution
are masked out because of the presence of foreground stars. The bottom-right
panel includes an SDSS-$g$ band image of the galaxy, shown at the same scale.
The radial profile of the constraints on the dust properties of
M82 are shown in Fig.~\ref{fig:M82_Rv} ($R_V$); Fig.~\ref{fig:M82_B}
(NUV bump strength); and Fig.~\ref{fig:M82_Ebv} (colour excess). In
all figures the grey dots correspond to individual measurements, with
a typical error bar shown on the left-hand side. The median and
root-mean-square values of the data, binned at fixed number of data
points per bin, are shown as black solid dots with error
bars. Finally, the solid line without error bars show the result if
the calibration -- presented in Tab.~\ref{tab:corr} -- is not
applied. Note the significant trend with galacto-centric distance.
The central region of M82 -- which due to projection effects include
{\sl both} the starbursting region and the outer areas -- has a more
standard attenuation law, compatible wih the $R_V\!=\!3.1$ and
$B\!=\!1$ values corresponding to the
\citet{Fitz:99} case. As we probe further out, the
total-to-selective extinction ratio parameter, $R_V$, decreases,
and so does the NUV bump strength, although we never find 
measurements compatible with a weak or non-existent bump.
The colour excess also depends on galacto-centric
distance, although in this case the correlation is more
tightly linked to the variation in the metallicity and star
formation distribution.

We note that this result is robust against a possible bias caused by
the inherent radial trend in age and metallicity found in
galaxies. The previous section explored through simulations the
possibility of a bias in the derivation of the parameters related to
the dust composition, to find no significant systematic trends
with age, metallicity or the different functional form of the star
formation history. Therefore, these trends are intrinsic radial
variations of the effective dust attenuation law ($R_V$,$B$), 
and the amount of dust ($E(B-V)$).

%%%%%%%%%%%%%%%%   Figure 8  %%%%%%%%%%%%%%%%%%%
%%%%%%%%%%%%%%%%%%%%%%%%%%%%%%%%%%%%%%%%%%%%%%%%
%%%%%%%%%%%%%%%%%%%%%%%%%%%%%%%%%%%%%%%%%%%%%%%%

Fig.~\ref{fig:M82_ParamCorr} shows the correlations among the three
dust-related parameters (using the same symbols as in the previous
figures). The range of values found {\sl within} M82 is
comparable with the average estimates of \citet{CSB:10} for 
a general sample of galaxies ($R_V\!=\!2.0$; $B\!=\!0.8$). Our
values are also consistent with the sample of nearby galaxies of 
\citet{Burg:05}, who find a range of values for the NUV bump 
strength, weaker than Milky Way standard, in the range\footnote{We
translate their $A_{\rm bump}$ to the $B$ parameterisation
of \citet{CSB:10} by a simple comparison of the effective attenuation
laws.} $B\!\sim\!0.5-0.6$.  Note the significant trend between $R_V$
and $B$ (top-left panel of Fig.~\ref{fig:M82_ParamCorr}), suggestive
of a strong link between the overall dust grain size distribution
(larger grains causing higher values of $R_V$) and the carrier of the
NUV bump strength. Although changes in the attenuation law can be
produced by changes in the age-dependent extinction of the geometry of
the stellar/dust components, our study -- based on a single galaxy --
would suggest that most of the variations are caused by the intrinsic
extinction law. A similar trend is found between $B$ and colour
reddening, although in this case the data mostly show that
regions with high reddening ($E(B-V)>0.5$) have high NUV bump
strengths (whereas less dusty regions can have a wide range of 
$B$, including $B\!=\!1$). In a similar way, regions with high
reddening are associated with higher values of $R_V$ (top-right
panel), corresponding to the presence of larger dust grains.

%%%%%%%%%%%%%%%%%%%%%%%%%%%%%%%%%%%%%%
%%%%%%%%%%   TABLE 2   %%%%%%%%%%%%%%%
%%%%%%%%%%%%%%%%%%%%%%%%%%%%%%%%%%%%%%
\begin{table}
\begin{center}
\caption{Corrections to the data, derived from simulations (see \S\ref{Sec:method}).
Col.~1 gives the parameter ($\pi$), cols.~2 and 3 are the slope and
intercept of the correction; col.4 gives the scatter with respect to
the full set of 3$\times$1,000 simulated data points (given as an
RMS); col.~5 is the bias, i.e.  the offset between the fit and the
data points.}
\label{tab:corr}
\begin{tabular}{lccc}
\hline
\multicolumn{4}{c}{$\pi_{\rm TRUE} = \alpha\pi_{\rm OBS}+\beta$}\\
\hline
$\pi$ & $\alpha$ & $\beta$ & RMS\\
\hline
$R_V$    & $1.030203$ & $-0.292036$ & $0.625$\\ 
$B$      & $1.032269$ & $-0.111693$ & $0.225$\\
$E(B-V)$ & $1.027146$ & $-0.104647$ & $0.194$\\
\hline
\end{tabular}
\end{center}
\end{table}
%%%%%%%%%%%%%%%%%%%%%%%%%%%%%%%%%%%%%%%%%%%%%%%%

%%%%%%%%%%%%%%%%%%%%%%%%%%%%%%%%%%%%%%
%%%%%%%%%%   TABLE 3   %%%%%%%%%%%%%%%
%%%%%%%%%%%%%%%%%%%%%%%%%%%%%%%%%%%%%%
\begin{table*}
\caption{Dust-related parameters and SSP ages derived from the analysis. Cols. 1 and 2
give the celestial coordinates of the central position of each
aperture. Col. 3 is the corresponding projected galactocentric
distance.  Col. 4 is the total-to-selective attenuation ratio. Col. 5
is the NUV bump strength.  Col. 6 is the colour excess, and col. 7
gives the base-10 logarithm of the SSP age, expressed in
Gyr. Uncertainties are quoted at the 1\,$\sigma$ level.}
\label{tab:results}
\begin{center}
\begin{tabular}{ccccccc}
\hline
RA &  Dec. &  $\Delta R$ &  Rv   &  B  & $E(B-V)$ & log Age\\
\multicolumn{2}{c}{J2000} & kpc & & & & Gyr\\
\hline
\vdots & \vdots & \vdots & \vdots & \vdots & \vdots & \vdots\\
$148.95876$ & $69.67534$ & $0.365$ & $2.594\pm 0.089$ & $0.936\pm 0.128$ & $1.025\pm 0.042$ & $-1.979\pm 0.060$\\
$148.95539$ & $69.67785$ & $0.195$ & $2.612\pm 0.095$ & $0.876\pm 0.135$ & $1.061\pm 0.048$ & $-1.921\pm 0.153$\\
$148.95203$ & $69.68039$ & $0.053$ & $3.626\pm 0.657$ & $0.793\pm 0.147$ & $0.743\pm 0.188$ & $-0.554\pm 0.710$\\
$148.94867$ & $69.68293$ & $0.171$ & $3.247\pm 0.570$ & $0.636\pm 0.161$ & $0.513\pm 0.186$ & $-0.147\pm 0.603$\\
$148.94530$ & $69.68547$ & $0.344$ & $2.777\pm 0.676$ & $0.537\pm 0.150$ & $0.386\pm 0.208$ & $-0.133\pm 0.703$\\
\vdots & \vdots & \vdots & \vdots & \vdots & \vdots & \vdots\\
\hline
\end{tabular}
\end{center}
This is an excerpt from the central region. The full version of this table can be retrieved online.
\end{table*}
%%%%%%%%%%%%%%%%%%%%%%%%%%%%%%%%%%%%%%%%%%%%%%%%

\subsection{The ISM around SN2014J in the pre-supernova phase}

Our data are especially relevant to explore the region around the
recent SN2014J event. We use the position of the supernova
from \citet{Tendulkar:14} to locate the relevant apertures in our
dataset. In Fig.~\ref{fig:M82_Rv}, \ref{fig:M82_B},
\ref{fig:M82_Ebv}, we show with red star symbols the estimates of
$R_V$,$B$,$E(B-V)$, respectively, measured in the apertures located within
30\,arcsec of the SN2014J event, corresponding to a projected distance
of 480\,pc. Notice that those regions feature higher values of
$R_V$($\sim$3-4) than the recent estimates determined
post-SN2014J \citep[$R_V\!=\!1.4\pm 0.1$,][orange
cross]{Aman:14}. In general, photometric studies of the
environment of type Ia supernova favour low values 
of $R_V$ \citep{Burns:14}, however the dispersion is quite
large \citep{Aman:15}.
For this particular case, even within the error bars and the dispersion of the
data, it is not possible to assume that post-SN2014J the dust
properties in this region are compatible. Therefore, our results
indicate that pre-SN2014J, this region had a dust content in the ISM
similar to the Milky Way standard. Such a result implies the
post-SN2014J observations should be explained either by a combination both
from dust in the ISM surrounding the supernova and the intrinsic
contribution from a circumstellar disk (\citealt{Foley:14}, although
see \citealt{Phillips:13}), or by
a significant change in the ISM towards the observed region.
Alternatively, one could also expect that supernovae have 
intrinsic variations in colour that are not accounted for, or
that the distribution of dust within the line of sight 
towards the supernova can produce the mismatch.

\subsection{Are the dust properties of M82 different from other galaxies?}

We note that the trends presented in this paper are at odds with
some recent results targeting our galaxy and M31.
In \citet{Nataf:13} and \citet{Nishiyama:09}, the dust attenuation
properties of the Milky Way are probed towards the Galactic bulge,
finding a steeper law -- i.e. lower $R_V$. In addition, \cite{Dong:14}
find similar trends in the Andromeda Galaxy. We note that the
populations towards the centre of M82 are significantly different from
these two galaxies.  The presence of a strong starburst will affect
the properties of the ISM, and the dust. In addition, the nearly
edge-on orientation of M82 along with its dusty ISM results in a complex
interpretation of the results towards small {\sl projected} 
galactocentric radii. 

Another interesting issue is the strong correlation between $R_V$ and
bump strength (top-left panel of Fig.~\ref{fig:M82_ParamCorr}).  A
recent study of the NUV bump strength in high redshift
galaxies \citep{KriekCon:13} suggests the opposite trend, namely a {\sl
decreasing} bump strength with increasing $R_V$. In contrast, detailed
studies in our Galaxy do not seem to find a correlation \citep{FM:07},
so there is still no clear indication whether such a correlation
should be expected in general.

We emphasize here that our methodology is unlikely to produce a
spurious correlation between these two parameters. The simulations
show that the (uncorrelated) dust parameters are retrieved without any
significant bias or artificial correlation among parameters. In
addition, we note that our study mainly differs
from \citet{KriekCon:13} in the sense that we explore {\sl internal}
variations within the same galaxy, rather than trends among galaxies.

%%%%%%%%%%%%%%%%%%%%%%%%%%%%%%%%%%%%%%%%%%%%%%%%
\section{Conclusions}
\label{Sec:Conc}

We revisit the NUV and optical photometry of M82, presented in HF14,
to constrain in more detail the properties of the dust attenuation
law.  Through the analysis of simulated data, we prove that it is
possible to factor out the contribution from the underlying stellar
populations, setting constraints on the NUV bump strength and the
total-to-selective attenuation ratio $R_V$. We find a significant
gradient with projected galactocentric distance
(Fig.~\ref{fig:M82_Rv}), with the central regions showing a standard
Milky Way attenuation law, progressing outwards towards an extinction
with a weaker NUV bump and steeper $R_V$, more characteristic of a
smaller grain size distribution, effectively tending towards the
standard attenuation law of the SMC \citep{Pei:92}.  We should note
here that our results correspond to a one-zone model, where the dust
only contributes as a foreground screen. Changes in the geometry of
the dust distribution can therefore cause variations on the dust
parameters presented here \citep[see, e.g.,][]{WG:00}.

%%%%%%%%%%%%%%%%   Figure 9  %%%%%%%%%%%%%%%%%%%
%%%%%%%%%%%%%%%%%%%%%%%%%%%%%%%%%%%%%%%%%%%%%%%%
\begin{figure}
\begin{center}
\includegraphics[width=8.5cm]{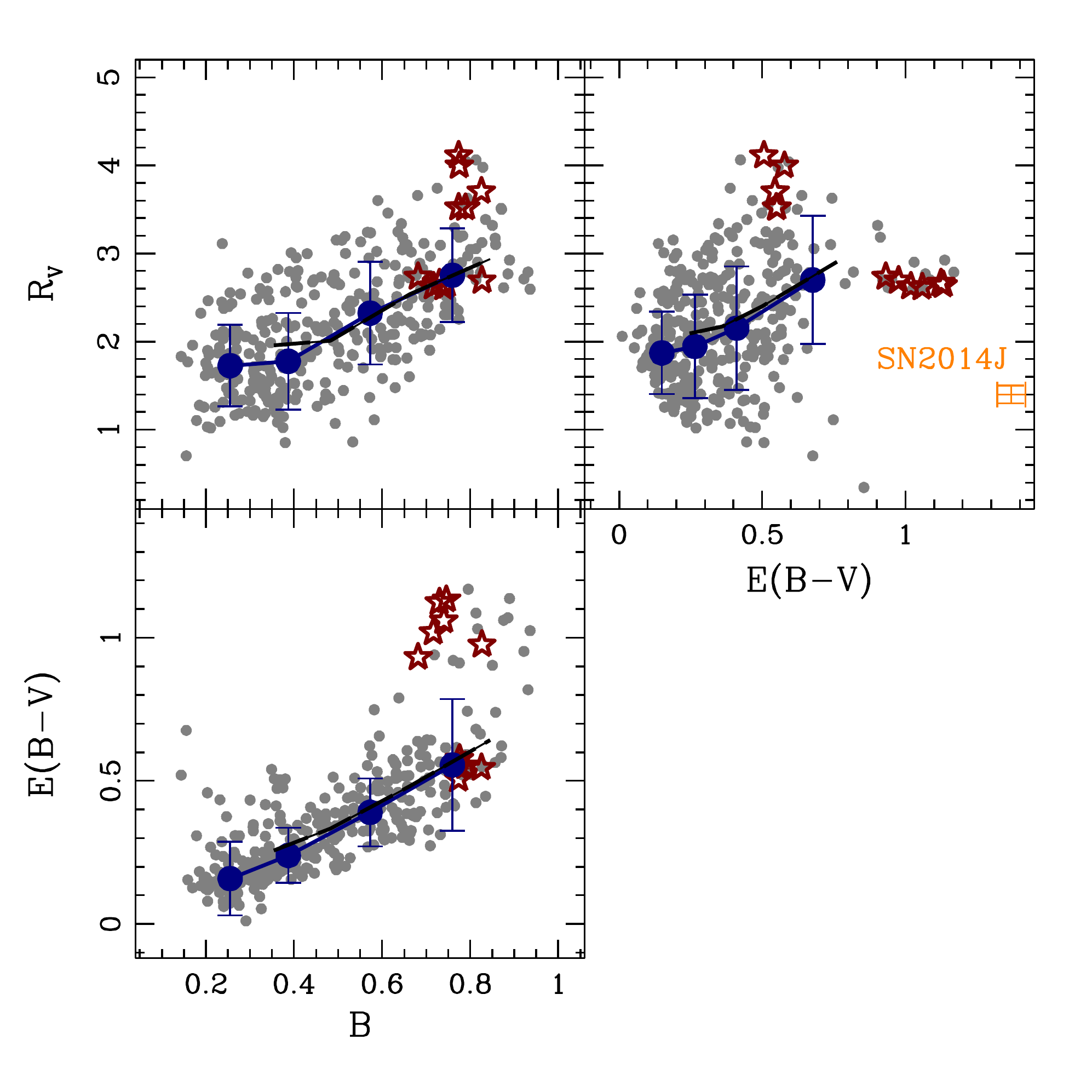}
\end{center}
\caption{Correlations among the fitted dust-related parameters.
The symbols are the same as in Fig.~\ref{fig:M82_Rv}.}
\label{fig:M82_ParamCorr}
\end{figure}
%%%%%%%%%%%%%%%%%%%%%%%%%%%%%%%%%%%%%%%%%%%%%%%%

As regards to the NUV bump strength, we find a systematic decrease
outwards. The weaker bump in the outer regions can be indicative of
changes in the dust properties \citep{GCW:97}, although an
age-dependent attenuation law can produce similar
effects \citep{Panuzzo:07}. However, our observational constraints go
in the opposite way with respect to the models. A trend towards a
weaker bump is found at large galacto-centric distances, {\sl away}
from the central starburst, where the dust content is lower, and where
$R_V$ is low, indicative of smaller dust grains (see
Fig.~\ref{fig:M82_ParamCorr}). Our results confirm HF14 in the sense
that a Milky Way law, with a significant NUV bump is favoured with
respect to a \citet{Calz:01} law, in the central starburst.

Our work is most relevant to the analysis of the environment of the
recent type Ia supernova SN2014J. Our data were taken before the
event, and therefore probe the properties of the dust of the ISM. At
the position of SN2014J, we find a standard dust component, typical of
the Milky Way, with $R_V\!\sim\!3$ and $B\!\sim\!1$. Therefore, the
recent measurements with lower values of $R_V$ reflect either the presence of
a circumstellar component, or a significant change in the dust properties
within the volume probed by the post-SN2014J measurements.

\section*{Acknowledgements}
We would like to thank the anonymous referee for very valuable comments and
suggestions.  This research has made use of data and software provided
by the High Energy Astrophysics Science Archive Research Center
(HEASARC), which is a service of the Astrophysics Science Division at
NASA/GSFC and the High Energy Astrophysics Division of the Smithsonian
Astrophysical Observatory. We acknowledge use of imaging data from the
Sloan Digital Sky Survey (SDSS). Funding for SDSS-III has been
provided by the Alfred P. Sloan Foundation, the Participating
Institutions, the National Science Foundation, and the U.S. Department
of Energy Office of Science. The SDSS-III web site is
http://www.sdss3.org/. We also thank the Mikulski Archive for Space
Telescopes (MAST). STScI is operated by the Association of
Universities for Research in Astronomy, Inc., under NASA contract
NAS5-26555. Support for MAST for non-HST data is provided by the NASA
Office of Space Science via grant NNX13AC07G and by other grants and
contracts.

%%%%%%%%%%%%%%%%%%%%%%%%%%%%%%%%%%%%%%%%%%%%%%%%
%%%%%%%%%%%%%%%   REFERENCES   %%%%%%%%%%%%%%%%%
%%%%%%%%%%%%%%%%%%%%%%%%%%%%%%%%%%%%%%%%%%%%%%%%

\label{lastpage}

\end{document}